\documentclass[manuscript]{aastex62}
\usepackage{graphbox}
\definecolor{black}{rgb}{0,0,0}
\definecolor{blue}{rgb}{0,0,1}
\definecolor{green}{rgb}{0,1,0}
\definecolor{red}{rgb}{1,0,0}
\definecolor{brown}{rgb}{0.4,0.2,0}
\definecolor{darkgreen}{rgb}{0,0.7,0}
\definecolor{darkblue}{rgb}{0.0,0.0,0.5}
\definecolor{red}{rgb}{1,0,0}
\definecolor{deepmagenta}{rgb}{0.8, 0.0, 0.8}

\begin{document}

\title{Efficient computation of collisional $\ell$-mixing rate coefficients in astrophysical plasmas}

\correspondingauthor{D. Vrinceanu}
\email{daniel.vrinceanu@tsu.edu}

\author{D. Vrinceanu}
\affiliation{Department of Physics, Texas Southern University, Houston, TX 77004, USA}

\author{R. Onofrio}
\affiliation{Dipartimento di Fisica e Astronomia ``Galileo Galilei,'' Universit\`a di Padova, Via Marzolo 8, 35131 Padova, Italy}
\affiliation{Department of Physics and Astronomy, Dartmouth College, 6127 Wilder Laboratory, Hanover, NH 03755, USA}

\author{J. B. R. Oonk}
\affiliation{Leiden Observatory, University of Leiden, P.O. Box 9513, 2300 RA Leiden, The Netherlands}
\affiliation{Netherlands Institute for Radio Astronomy (ASTRON), Oude Hoogeveensedijk 4, 7991 PD Dwingeloo, The Netherlands}
\affiliation{SURFsara, P.O. Box 94613, 1090 GP Amsterdam, The Netherlands}
  
\author{P. Salas}
\affiliation{Leiden Observatory, University of Leiden, P.O. Box 9513, 2300 RA Leiden, The Netherlands}

\author{H. R. Sadeghpour}
\affiliation {ITAMP, Harvard-Smithsonian Center for Astrophysics, Cambridge, MA 02138, USA}

\begin{abstract}
We present analytical expressions for direct evaluation of $\ell$-mixing rate coefficients in proton-excited hydrogen atom collisions and describe a software package for efficient numerical evaluation of the collisional rate coefficients. Comparisons between rate coefficients calculated with various levels of approximation are discussed, highlighting their range of validity. 
These rate coefficients are benchmarked via radio recombination lines for hydrogen, evaluating the corresponding departure coefficients from local thermal equilibrium.
\end{abstract}


\section{Introduction}
Energy-conserving angular momentum-changing $n\ell \rightarrow n\ell'$ transitions induced in collisions between Rydberg atoms and low velocity ions are needed for accurate comparison between astrophysical observations and models which employ atomic theory for temperature and density diagnostics in diffuse atomic clouds, H II regions and various nebulae. With ion collision-induced angular momentum mixing rate coefficients scaling as $n^4$, values as large as a few times $10^{5}$ cm$^{3}$/s are possible for principal quantum numbers near $n \sim 200$. Accurate and efficiently calculated rate coefficients are hence necessary to interpret a host of astrophysical processes, such as radio recombination lines from hydrogen (HRRL) and carbon (CRRL) as tracers of the neutral phase of the interstellar medium (ISM) 
(e.g. \citealt{Oonk2015,Oonk2017,Salas2018}), and from hydrogen as a tracer of gas ionized by young stars (HII regions)
(e.g. \citealt{Roelfsema1992,Anderson2011}). The recombination of hydrogen and helium in the early Universe, and the primordial abundance of helium, are also examples of processes affected by collision physics \citep{Izotov2010,Chluba2006}.

In their pioneering work, Pengelly and Seaton obtained proton-Rydberg hydrogen collisional cross sections  for dipole allowed transitions within the Born-Bethe approximation \citep{Pengelly1964} (PS64 hereafter). Given that the probability for $\Delta \ell =\pm 1$ transition falls off asymptotically as the inverse square of the impact parameter and the cross section hence becomes logarithmically divergent, PS64 invoked a set of cutoff conditions to circumvent this divergence. The arbitrariness implicit in choosing these cutoff conditions was re-examined in \citet{Vrinceanu2001a,Vrinceanu2001b} (VF01a and VF01b), where a non-perturbative closed form solution for the transition probability was found. 

A semiclassical (SC) rate coefficient for arbitrary $\ell$-changing collisions was derived in \citep{Vrinceanu2012} {(VOS12)}, and was shown to be in agreement with both classical trajectory Monte Carlo simulations and numerically integrated quantum rate coefficients for transitions with $|\Delta\ell| > 1$. For the dipole allowed transitions ($|\Delta \ell =1|$), probabilities evaluated with VOS12 SC increase linearly with impact parameter and at a critical value abruptly vanish. This unphysical behavior of the SC transition probability, first noticed in \cite{Storey2015}, was addressed by the modified PS64 (PS-M) approximation \citep{Guzman2016,Guzman2017}.
A further improved SC probability with the correct asymptotic behavior, yielding a more accurate formula for the dipole transitions, was later derived by \citet{Vrinceanu2017} (VOS17).

In this work, we derive a computationally efficient formula of the rate coefficients for  collision-induced dipole transitions, accurate for a broad range of $n$, temperatures ($T$) and electron number densities ($n_e$). This is achieved by using a  closed-form expression for the dipole transition rate coefficients, therefore overcoming the need for the explicit calculation of cross sections. 
In addition, this formulation circumvents an unphysical behavior of the PS64 rate coefficients, which become negative for a range of astrophysically relevant parameters. The resulting rate coefficients are then used to evaluate the departure coefficients from statistical equilibrium of HRRL.

\section{Rate coefficients for angular momentum changing transitions in proton-Rydberg hydrogen scattering}

The rate coefficients for $n$-conserving, $\ell$-changing transitions are obtained by integrating the corresponding transition probability, $P_{n\ell \rightarrow n\ell'}$, over the impact parameter, $b$, and thermal distribution of the projectile velocity, $v$,
\begin{equation}
\label{thermal}
k(n, \ell,\ell', T, R_c) = 2 \pi \int_0^\infty f_{MB}(v)\; v dv \int_0^{R_c} P_{n\ell\rightarrow n\ell'}(b,v)\; b db \;,
\end{equation}
with $f_{\text{MB}}$ the Maxwell-Boltzmann distribution at temperature $T$. 
The cutoff distance $R_c$ is required to regularize the divergent integral for dipole allowed transitions ($|\Delta\ell| = 1$), when the transition probability decreases too slowly for $b\rightarrow\infty$. For all other cases, the integral is finite as $R_c\rightarrow\infty$. The transition probability does not depend on $b$ and $v$ independently, but through the collision parameter $\alpha = 3 Z n \hbar/(2 m v b)$ (VF01a), so that the double integral in Eq.~(\ref{thermal}) is reduced to
\begin{equation}\label{rate}
k(n, \ell,\ell', T, R_c) = 
n^4 a_0^2 v_0 \sqrt{\frac{8 \pi \mu v_0^2}{k_B T}} \int_0^\infty z P_{n\ell\rightarrow n\ell'}(z)\; e^{-\theta z^2/2}\;dz\;,
\end{equation}
where $a_0 = 5.29177\times 10^{-11}$ m is the Bohr radius, $v_0 = 2.18769 \times 10^6$ m/s is the atomic unit of velocity, and the integration variable is $z =  3/(2 n \alpha)$. The parameter $\theta$, small for large $R_c$, is defined as
\begin{equation}\label{theta}
\theta = n^4 \; \frac{\mu v_0^2}{k_B T} \frac{a_0^2}{R_c^2}\;, 
\end{equation}
where $\mu$ is the reduced mass of the projectile - target system.

When $R_c$ is chosen to be the Debye length $\lambda_D = \sqrt{\epsilon_0 k_B T/( n_e e^2)}$, as in PS64, this parameter becomes
\begin{equation}\label{theta2}
\theta = 1.704675 \times 10^{-10} n^4  \frac{n_e}{T^2}\;,
\end{equation}
with $T$ in $K$ and $n_e$ in cm$^{-3}$.
Depending on the specific physical situations, other choices for $R_c$ are possible as discussed in PS64, changing the $\theta$ parameter accordingly.

\begin{figure}[b]
\vskip0.1in
\centerline{\includegraphics[width=4.in, align=t]{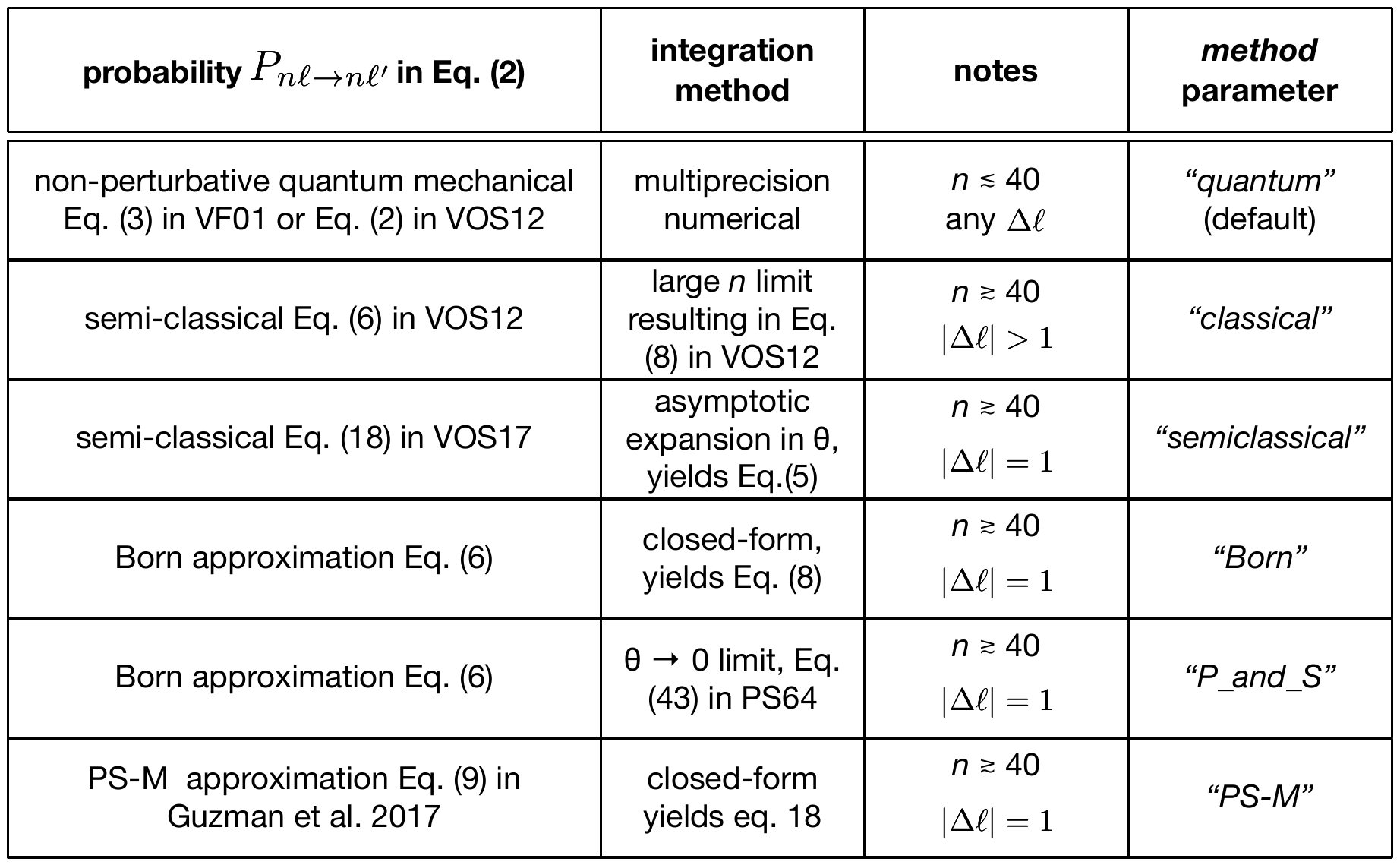} \hfill \includegraphics[width=3.in, align=t]{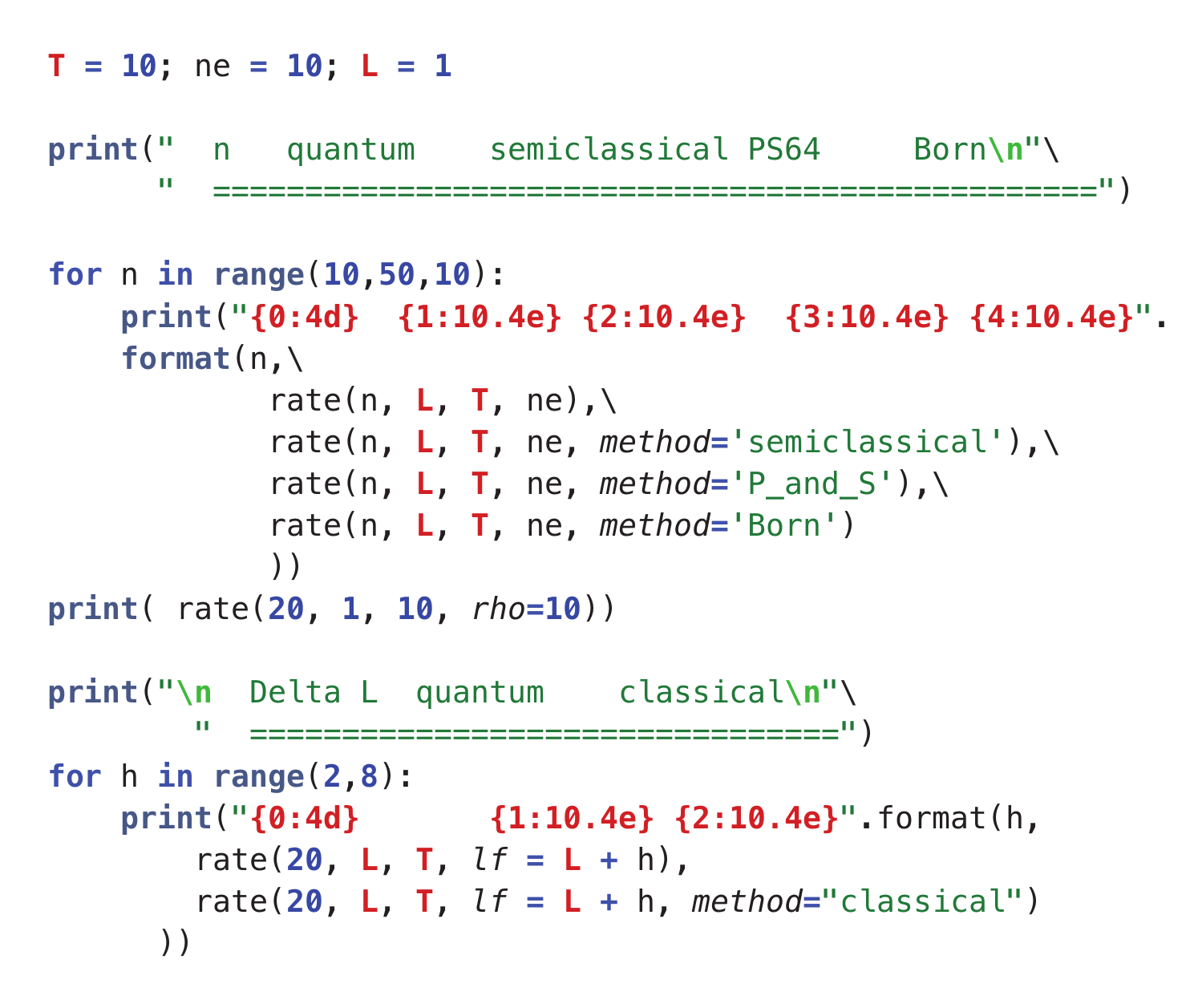}}
\caption{\label{fig1} The \texttt{Lmixing} package provides the rate coefficient $k(n,\ell,T, n_e)$ by using several approximations for the transition probability $P_{n\ell\rightarrow n\ell'}$ in Eq. (2), as outlined in the table on the left. An example code showing the usage of the package is displayed on the right. By default, the function \texttt{rate} calculates the rate coefficient for both dipole allowed transitions $\ell' = \ell \pm 1$ by integrating numerically the quantum mechanical probability. For $n \gtrsim 40$, when integration requires extended precision and takes a longer time to complete, faster, but less accurate, approximations can be selected with the optional parameter {\em method}.
}
\end{figure}

Equation~(\ref{rate}) is the starting point for the calculation of the rate coefficient using various approaches depending on the choice of $P_{n\ell \rightarrow n\ell'}$. The relationships among these approximations reflect the organization of the software package \citep{Vrinceanu2018}, and are illustrated in the diagram in Fig. \ref{fig1}, and discussed in detail below. 

The rate coefficients are calculated by numerically integrating Eq. (\ref{rate}). This integral does not pose difficulties for relatively low principal quantum numbers $n \lesssim 40 $ when using the VF01b non-perturbative quantum mechanical $P_{n\ell \rightarrow n\ell'}$. However, for larger $n$, the calculation does not converge because of truncation errors and near cancellation of large terms. These problems are addressed in our code by using exact number arithmetic for large factorials and extended floating point precision for large order polynomials. 
The computational time increases as $\sim \ell \times n$, to the point that accurate calculations become impractical; for example, the calculation for $n=1000$ can take up to 5 minutes. More efficient, but less accurate approximations are implemented in the code for large $n$ and $\ell$, when numerical integration of quantum probability is slow.
The {\it method} parameter in the code selects the desired procedure, either quantum, or from the menu of approximations discussed below, and should be chosen based on the accepted level in the trade-off of accuracy versus computing cost.  

{VOS12} SC formula works well for moderate to large  $|\Delta\ell|$, but loses accuracy for larger $|\Delta\ell|$, for which the rate coefficients are small and may be neglected. The {VOS12} SC formula has been derived by integrating the classical limit of {VF01a} transition probability, and has been validated by extensive classical trajectory Monte Carlo simulations. The SC approximation fails for dipole  transitions $|\Delta\ell| = 1$, because the SC transition probability increases linearly with the impact parameter and drops abruptly after a critical value, instead of decreasing as $b^{-2}$, as obtained with perturbation theory (PS64).

For dipole transitions, the code provides rate coefficients for combined $\ell\rightarrow\ell+1$ and $\ell\rightarrow\ell-1$ transitions. In the Born approximation, $P_{n\ell \rightarrow n\ell'}$ is assumed to be 1/2 for $R<R_1$, after which it decreases as $1/b^2$. The PS64 formula is derived from this approximation by adopting the additional assumption that $R_1 < R_c$. This assumption fails for large $n_e$ and low $T$, limiting the range of applicability of PS64 to $n^2 (n^2 - \ell^2 - \ell -1) n_e /T^2 < 2.98\times 10^9$ cm${}^{-3}$/K${}^2$. Within this range of parameters, PS64 is in reasonable agreement with the non-perturbative quantum results, as demonstrated in the next section. Outside this range, PS64 rate coefficients become negative {\citep{Salgado2017,Guzman2017}.}

The PS-M approximation introduced in \citet{Guzman2017} replaces the constant 1/2 transition probability in the PS64 model with a linearly increasing one, and obtains the rate coefficient by averaging the resulting cross section over all energies, including those neglected by PS64 for $R_1 < R_c$. The approximate PS-M rate coefficients shown in table I in \citet{Guzman2017} are positive even when PS64 are negative, and yield an overall better agreement with the quantum results, although significant deviations are noted in some cases, up to a factor of 10, probably due to the simplicity of the model.

The shortcomings of the VOS12 SC  approximation for dipole allowed transitions were addressed in VOS17 by deriving a more accurate SC probability with the correct large $b$ asymptotics.
When used in Eq.~(\ref{rate}), the VOS17 transition probability leads to a SC rate coefficient for dipole allowed transitions 
\begin{equation}\label{SC-rate}
k(n, \ell, T, n_e) = \sqrt{\frac\pi 2} a_0^2 v_0 \sqrt{\frac{\mu v_0^2}{k_B T}} D_{n\ell} 
\left[
\frac{3 \sqrt{\pi}}{4 x^{3/2}}\; \text{erf}(\eta \sqrt{x}) - \frac{3\eta}{2 x} e^{-\eta^2 x} + 
\sum_{k=0}^{N_t} A_k (B_k - 3 \gamma - \log(4 x)) x^k 
\right]
\end{equation}
where erf is the error function, $D_{n\ell} = 6 n^2 (n^2 - \ell^2 - \ell - 1)$, $\eta = 0.277855$  is the solution of the equation $j_1(1/z)^2 = z/6$, with $j_k(u)$ a spherical Bessel function of order $k$ and argument $u$, and  $x =  3 D_{n\ell} \theta /(4 n^4)   = n_e  D_{n\ell}/ (7.82162\times 10^9 T^2)$.

The coefficients in the asymptotic expansion can be calculated up to the truncation order $N_t$ by using
\[
A_k = \frac{18 \times 4^k}{(k + 2)(k + 3)(2k + 3) \; k!\; (2k + 1)!}
\]
and 
\[
B_k = \frac 1{k + 2} + \frac 1{k + 3} + \frac{2}{2k + 3} + H_k + 2H_{2k + 1}
 \]
where $H_k = \sum_{j=1}^k 1/j$ for $k>0$, $H_0=0$, with $k$ the $k$-th harmonic number and $\gamma = 0.57721$ the Euler constant.
A derivation of this formula, and a list of the first eleven $A_k$ and $B_k$ coefficients, is given in the appendix. Our experience shows that a truncation order $N_t \sim 16 - 20$ is sufficient to provide accurate results.

\section{Results}

As illustrated in Fig. (1), the Python module \verb!Lmixing! \citep{Vrinceanu2018} has a function that calculates the angular momentum mixing rate coefficient. The required  arguments are the principal quantum number $n$, the initial angular momentum $\ell$, and the temperature in Kelvin. The optional arguments are the electron number density $n_e$ in cm$^{-3}$, the final angular momentum $\ell'$, and the method of calculation. 
The choices for this argument are: \verb!quantum! (default value) {using Eq. (2) with the probability defined by Eq. (3) in VF01b}, \verb!semiclassical! {using Eq. (5)}, \verb!Born! for the Born approximation {Eq. (8)}, \verb!PS-M! for the approximation in Eq. (\ref{PS-M}) originally introduced in \citet{Guzman2017}, and \verb!P_and_S! which implements the PS64 method. If $\ell'$ is not provided, the rate for the combined $\ell \rightarrow \ell\pm 1$ transitions is calculated. If $\ell'$ is given, one can then choose the exact quantum calculation, or the SC approximation {(Eq. (8) in VOS12)} with the value for the method parameter \verb!classical!, provided that $|\Delta\ell| > 1$.

\begin{figure}[t]
\includegraphics[width=3.5in]{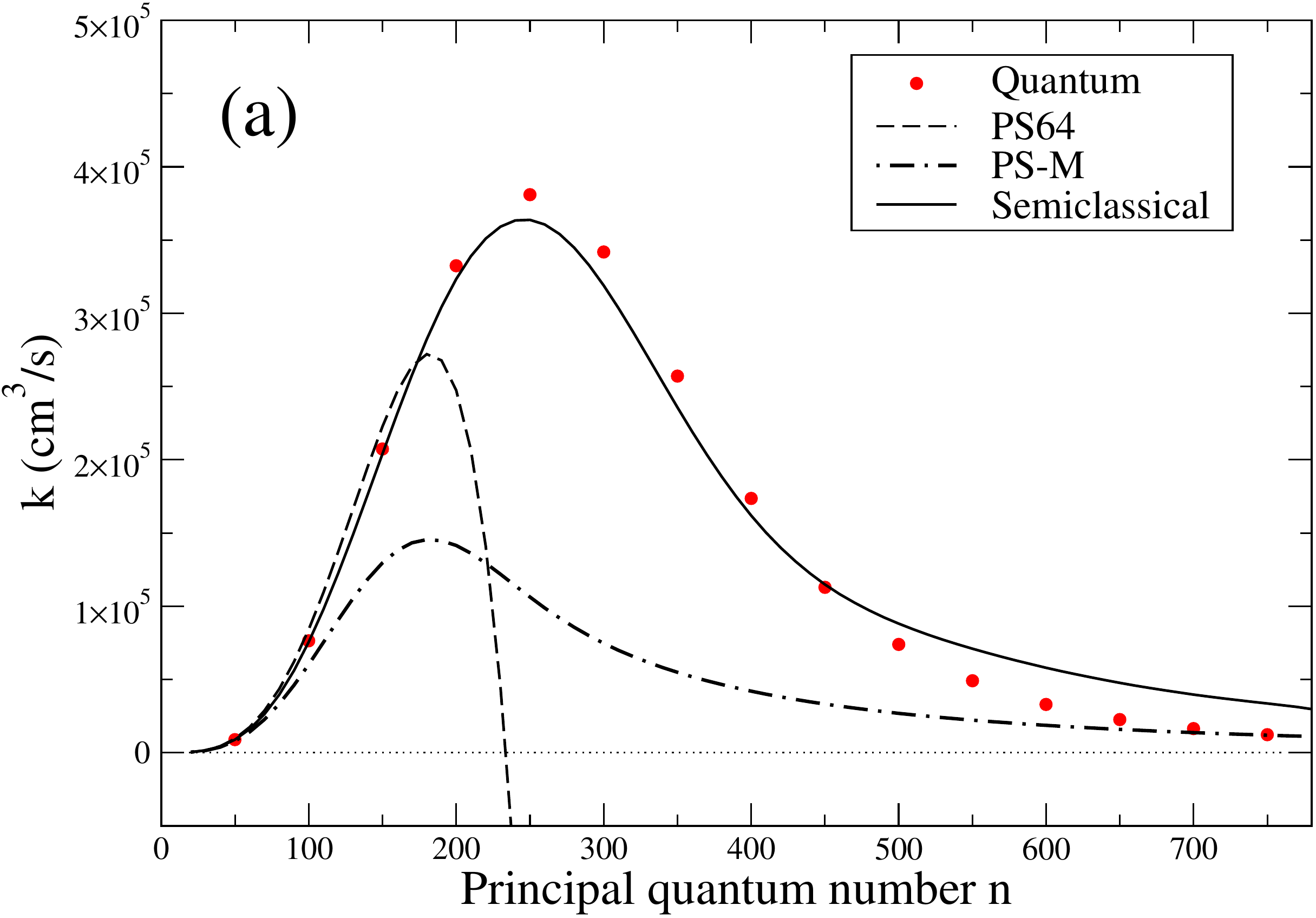}
\includegraphics[width=3.5in]{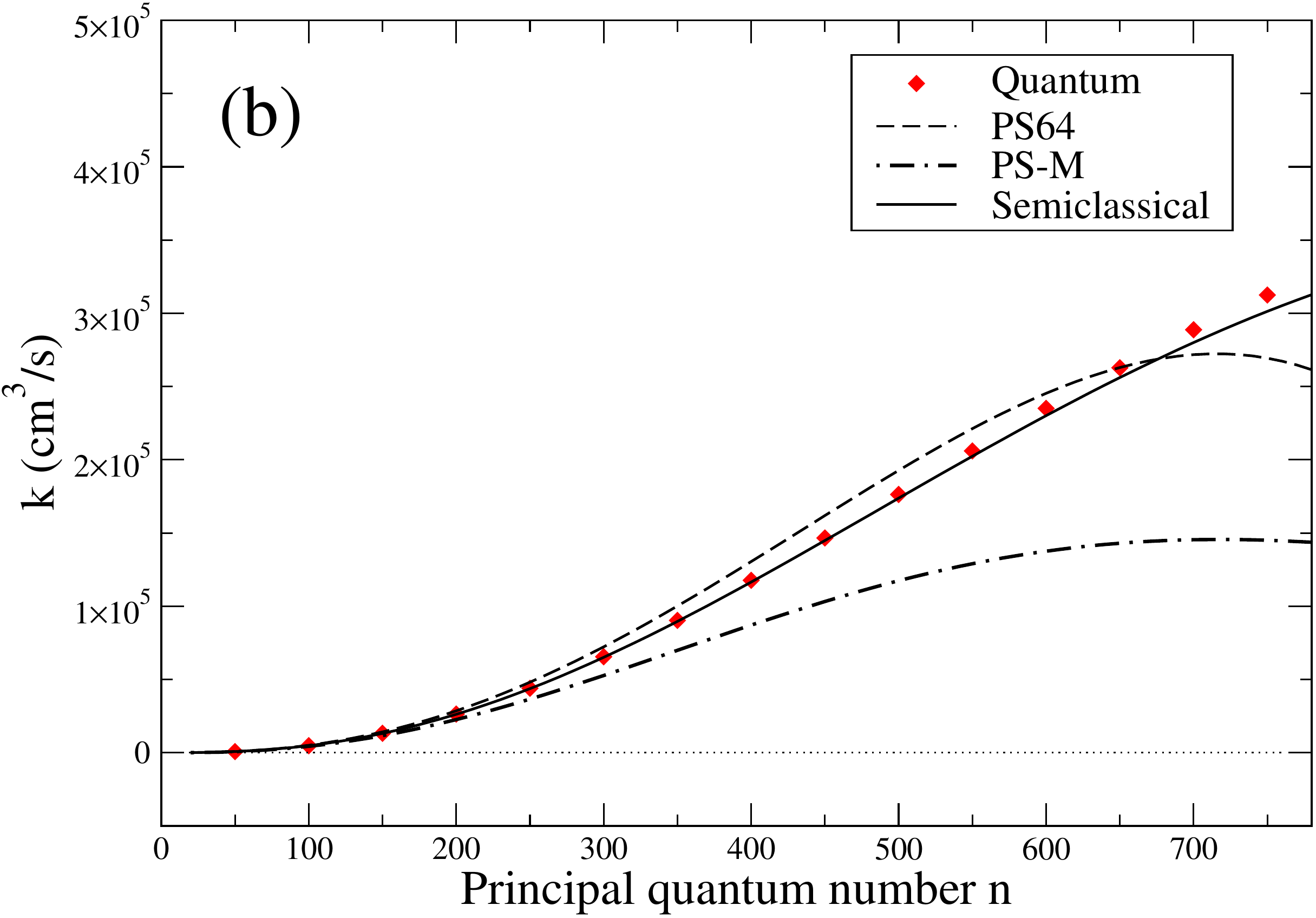}
\caption{\label{fig2}
Rate coefficients for combined $(n,\ell)\rightarrow (n,\ell\pm 1)$ transitions versus $n$ for transitions 
of extreme eccentricities $\ell=1$ (a) and $\ell=n-2$ (b), at  $T=10$ K and $n_e=100$ cm$^{-3}$.  
The solid line is obtained by using Eq.~(\ref{SC-rate}) 
with $N_t=16$, dots refer to the direct integration of the quantum transition probabilities, dashed lines and dot dashed lines represent the PS64 Eq. (\ref{PS}) and PS-M Eq. (\ref{PS-M}) results, respectively.} 
\end{figure}

Figure \ref{fig2} shows a comparison between the rate coefficients coefficients for dipole allowed transitions for $\ell = 1$ and $\ell = n-2$ angular momenta, calculated by integrating the VF01b quantum formula, the PS64 approximation and the SC approximation in Eq.~(\ref{SC-rate}). The evaluation of the quantum case for extremely large $n$ is slow even when low accuracy results are sufficient, due to the use of multiprecision floating point arithmetic necessary to prevent truncation error and loss in precision. The calculations in Fig.~\ref{fig2} required 400 digits of precision and took several hours for $\ell=1$ cases and two days for $\ell = n-2$ cases to complete on a single processor.  At around $n = 200$, in the $\ell=1$ case, the PS64 approximation fails and becomes negative for higher $n$, as first discussed in \citet{Salgado2017}, and in the caption of Table 1 in \citet{Guzman2017}, while the SC approximation is in good agreement with the quantum results over a much larger range of $n$. For sufficiently low T and high $n_e$, the  SC rate coefficients for $n \gtrsim 500$ overestimate the corresponding quantum result, then underestimate the latter at even higher $n$ (not shown in Figure \ref{fig2}), eventually becoming negative. However, unlike the case of PS64, the negative-defined region occurs for progressively higher values of $n$ as more terms in the expansion of Eq.~(\ref{SC-rate}) are considered.
In the $\ell=n-2$ case, the PS64 results are more reliable with respect to the corresponding ones at low angular momentum, and begin to diverge from the exact results for $n > 700$, becoming negative at around $n = 1000$. The overall shape of the dependence of the rate coefficient, $k$ with $n$, is the result of two competing factors, as seen in Eq.~(\ref{SC-rate}): on the one hand the prefactor in front of the integral increases as $~n^4$, while  the integral decreases roughly as $1/\theta$. { For comparison, we also include in Figure \ref{fig2} PS-M rate coefficients which behave more consistently with $n$ than the PS64 model in the $\ell=1$ case, and approach the quantum results at large $n$. However, other than the large $n$ limit for 
$\ell=1$, PS-M predicts rate coefficients which are consistently smaller than QM and SC values, as particularly evident in the $\ell=n-2$ case.}

\begin{figure}
\centerline{\includegraphics[width=7.5in]{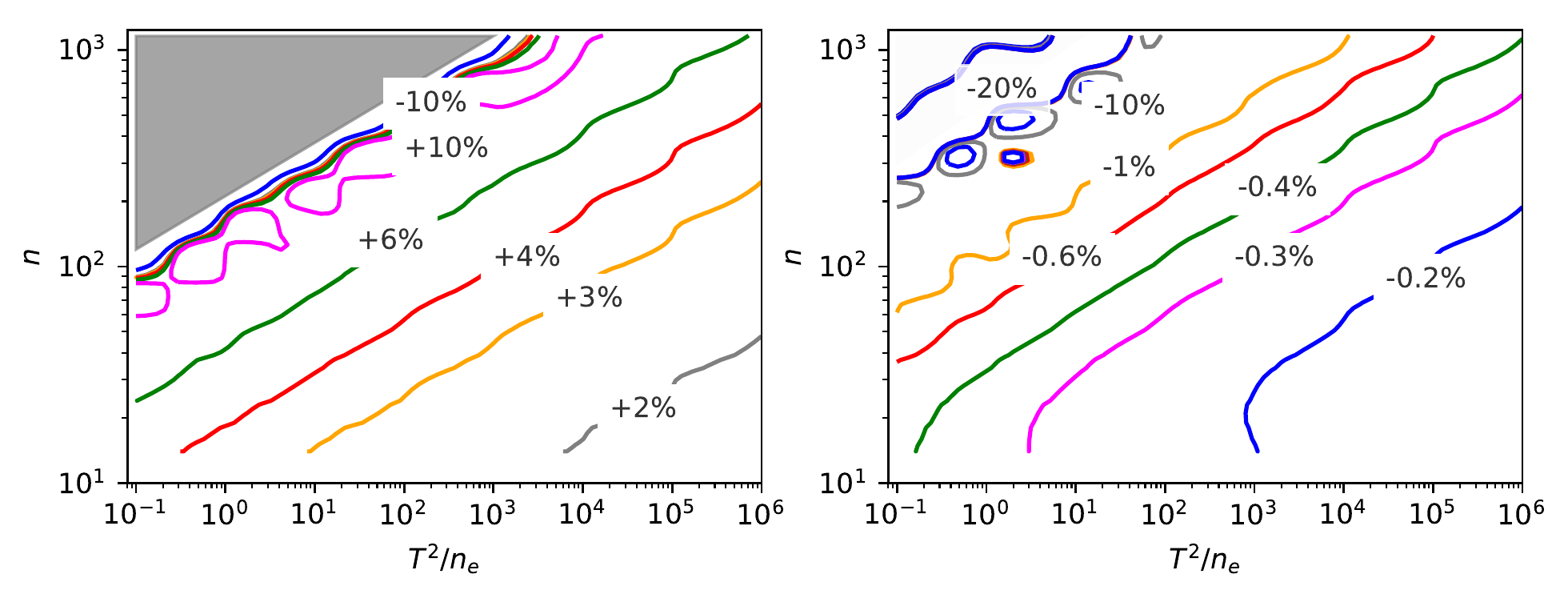}}
\caption{\label{fig3}
Discrepancies for the rate coefficients calculated with the PS64 approach (left)
and the SC approximation (right), both referred to the non-perturbative
rate coefficients evaluated using the quantum formula, for a broad range of $n$ and
$T^2/n_e$.}
\end{figure}

In Figure \ref{fig3}, the accuracy of PS64 approximation (left panel) and of the SC approximation (right panel), both with respect to the quantum rate coefficients, is shown for astrophysically relevant values of $n$, $T$ and $n_e$, and for the $\ell=1$ transition. The SC approximation is accurate within $1\%$ over a wide range of $T$ and $n$, while the accuracy of PS64 is roughly one order of magnitude worse at the same point in the $n-T^2/n_e$ plane. The PS64 approximation yields negative rate coefficients in the upper left corner of the diagram, {\it i.e.} in the low $T$, high $n_e$, and high $n$ regime. As already discussed for Figure \ref{fig2}, the SC rate coefficients  also become negative in the upper left corner, but in comparison to the PS64 case this occurs for a smaller region with size inversely proportional to the number  of terms $N_t$ in Eq.~(\ref{SC-rate}). These cases are extreme, but important for the interpretation of HRRLs which probe the low electron density, cool ISM (e.g. \citet{Salgado2017} and references therein).
The SC approximation obviously becomes less accurate in the region for which the rate coefficients approach negative values. This is also expected as for these parameters $\lambda_D \leq n^2 a_0$, indicating that the binary collision assumption may fail, with the $\ell$-mixing instead ruled by many-body interactions. If the opposite case of $\ell=n-2$ is considered, the rate coefficients for PS64 and the SC approximations are much closer, in line with what is expected by inspecting the corresponding curves in Fig. \ref{fig2}. Figure \ref{fig3}  puts on a more quantitative standing the recent debate on the accuracy of various proposed rate coefficients  as reported in  \citet{Storey2015,Guzman2016,Guzman2017,Williams2017}. 
The SC approximation is in general more accurate than the above approximations because the VOS17 transition probability agrees better with the quantum results. It was noticed in \citet{Guzman2017} that for some extreme cases, the PS-M overestimates the quantum results by a factor of 10.

\begin{figure}[t]
\centerline{
\includegraphics[width=6.in]{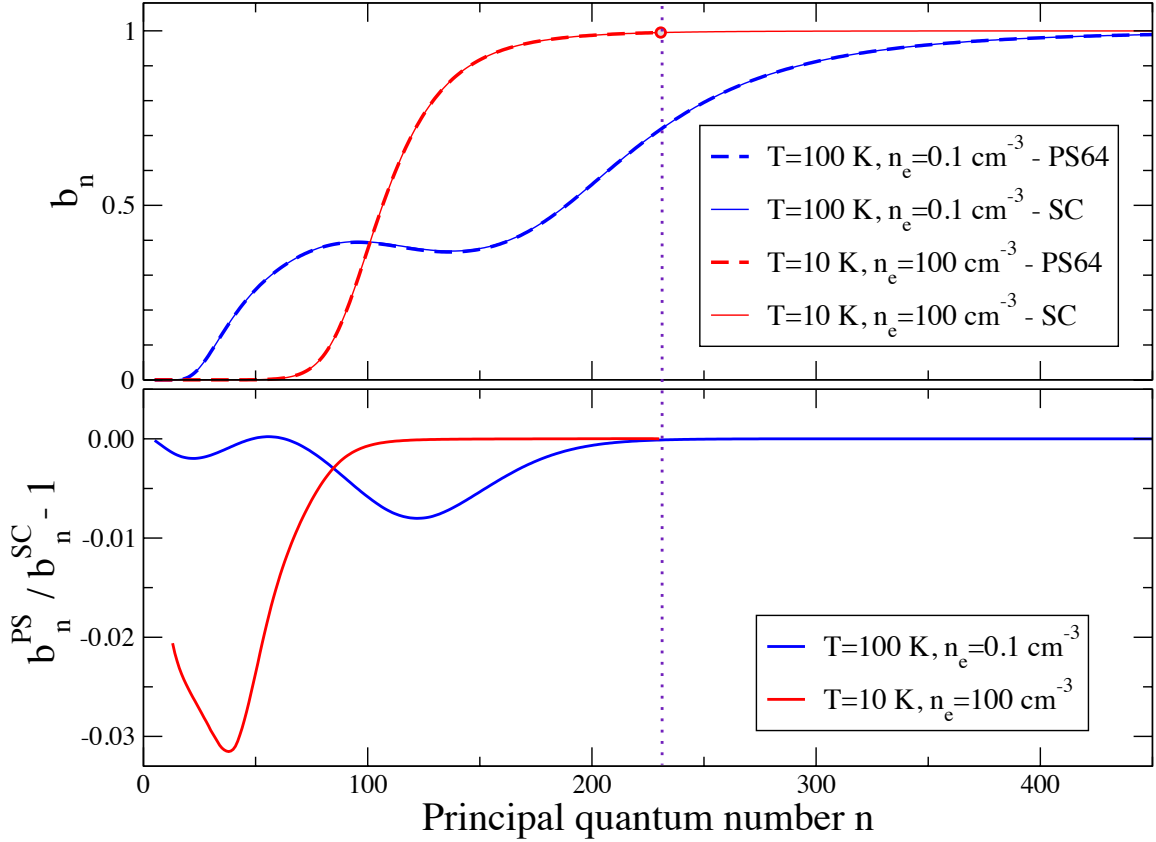}
}
\caption{\label{fig4}
Calculations of the $b_n$ departure coefficient from collisional-radiative simulations of hydrogen for
two cases: $T = 100$ K and $n_e = 0.1$ cm$^{-3}$ (blue lines), and $T = 10$ K and $n_e = 100$ cm$^{-3}$ (red lines). The upper plot shows $b_n$ calculated by using PS64 (dashed lines) and by using the SC approximation (solid lines). The lower plot shows the relative difference between the results using the two approximations, for each case considered. The dotted line and the circled point mark the maximum $n$ for which PS64 provides convergent results in the low $T$, high $n_e$ case. When converged, the PS64 and SC results agree with a maximum difference of few percent.}
\end{figure}
The HRRL results shown in Fig. 4 were produced with the Salgado et al. (2017a) models using the updated $\ell$-changing collision rate coefficients, Eq.~(\ref{SC-rate}) with $N_t=10$ and, as a comparison, also using the PS64 rate coefficients. The latter are known to be in good agreement with the quantum mechanical rate coefficients for sufficiently high $T$, high $n_e$ and low $n$.

We show in Fig.~\ref{fig4} the results for the departure coefficients from thermal population, $b_n$, for a homogeneous, one dimensional gas slab with: (i) $T_e$ = 100 K and $n_e$ = 0.1 cm$^{-3}$, and (ii) $T_e$ = 10 K and $n_e$ = 100 cm$^{-3}$. The former is a typical cool ISM case and the latter is an an extreme case. Both are exposed to a Galactic power-law radiation field $T_R \propto \lambda^{2.6}$ that is normalized at 100 MHz by $T_{R,100} = 2000$ K \citep{Salgado2017}. We find that the different $\ell$-changing collision rate coefficients primarily affect $b_n$ at low to intermediate $n$ values ($n \lesssim 300$) with differences up to a few percent for typical cool ISM conditions. As explained in \citet{Salgado2017}, but see also \citet{Hummer1987}, this is because at these intermediate $n$ levels collisions compete with spontaneous decay, effectively storing electrons in high $\ell$ sublevels for which radiative decay is less important. Notably, the $\ell$-changing collision rate coefficients presented here allow us to efficiently calculate $b_n$ values for high $n$, where the PS64 rate coefficients no longer apply, and which are important to studies of cool, partially ionized ISM (e.g. \citealt{Oonk2017, Salas2018}).

The HRRL optical depths are calculated using the product of $b_n \beta_n$, where the correction factor for stimulated emission $\beta_n$ can be seen as the derivative of $b_n$ \citep{Salgado2017}, such that small changes in $b_n$ can lead to somewhat larger changes in HRRL optical depth. These changes are measurable, but require very high signal to noise observations across a broad frequency range (i.e. 240-2000 MHz). Most HRRL observations have difficulties achieving such accuracy, and hence the calculated differences will be within current observational uncertainties for typical ISM conditions. We have also compared the HRRL results presented here with those computed using the VOS12 SC rate coefficients \citep{Salgado2017}. We find that for these cases the results agree to within a few percent for $n\gtrsim 300$. For lower $n$-values the agreement is less good.

We are currently implementing our new $\ell$-changing rate coefficients for CRRLs, as will be presented in a future work.  
Although we anticipate that the results may be qualitatively similar to those for HRRLs, from a quantitative standpoint we expect to have stronger dependence of the departure coefficients on the $\ell$-changing collision rate coefficients than HRRLs \citep{Salgado2017}.

\section{Conclusions}

We have introduced an efficient and accurate SC approximation
for $\ell$-mixing processes, allowing direct evaluation of collisional rate coefficients for a broad range of $n$, $T$, and $n_e$ relevant to astrophysical plasmas. We provide a Python module for evaluating these rate coefficients for easy integration in large-scale radiative-collisional simulation codes. The Python code also implements a multi-precision numerical integration of the quantum rate constants for small to moderate $n$. The relationship and range of validity and accuracy for various schemes to evaluate $\ell$-changing rate coefficients are elucidated. Furthermore, we identify the range of parameters for which PS64 rate coefficients become unphysical.
Efficient computer codes for the evaluation of $\ell$ mixing rate coefficients with various levels of approximation are beneficial for accurate astrophysical modeling. As noted in the recent release of {\em Cloudy} \citep{Ferland2017}, different choices for the rate coefficients can lead to differences up to
10\% in the predicted line intensities, which are larger than the precision of current observations.
The rate coefficients proposed here are in better agreement with the more rigorous, but computationally less efficient quantum rate coefficients.

\section*{Acknowledgments}

This work was supported by the National Science Foundation through a grant to ITAMP at the Harvard-Smithsonian Center for Astrophysics. One of the authors (DV) is also grateful for the support received from the National Science Foundation through grants PHY-1831977 and HRD-1829184. J. B. R. O. and P. S. acknowledge financial support from the Dutch Science Organization (NWO) through TOP grant 614.001.351.

\software{Lmixing \citep{Vrinceanu2018}}

\section{Appendix}

This section presents the derivations of the main results and implementation details of the computational module. According to the organization of the code illustrated in Fig.~\ref{fig1} we will discuss the four main sections: the integration of the exact quantum formula, the Born approximation, the classical approximation and the semiclassical approximation. These approaches are derived from Eq.~(\ref{rate}) either by keeping the exact, but computationally expensive form for the transition probability $P_{n\ell\rightarrow n\ell'}$, or by replacing it by an approximate expression that provides quicker results with a limited range of validity. The goal is to calculate the {dimensionless} integral {in Eq.(\ref{rate})} 
\begin{equation}\label{integral}
\tilde k(n,\ell,\ell',\theta) = \int_0^\infty z P_{n\ell\rightarrow n\ell'}(z) e^{-\theta z^2/2}\; dz.
\end{equation}

The exact calculation uses the quantum transition probability derived in (VF01b), that has practical use limited to small ($n < 30$) quantum numbers in regular computer arithmetic. The main two reasons for this difficulty are the combinatorial Wigner 6-j symbols involving factorials of large integers that cannot be represented exactly with integer types, and the loss of precision in the calculation of polynomials of large order with alternating terms. Our code takes advantage of the unlimited size integer type in Python by pre-computing tables of large factorials, and by using fixed-point representations of real numbers with a prescribed, but arbitrary, number of decimal figures. As a rule of thumb, we find that a calculation for quantum numbers $n$ requires setting the precision at $n/2$ digits. The numerical integral Eq.~(\ref{integral}) is calculated with a recurrent Gauss-Lobatto-Kronrod algorithm from \cite{Press1992} adapted for the use of extended precision real numbers. 

A Born approximation is obtained from perturbation theory (VOS17) only for dipole allowed transitions. The cumulative probability for both $\ell\rightarrow\ell'$ transitions is approximated as
\begin{equation}
P^{(B)} = \frac 12 \left\{ \begin{array}{ll}
1            & \mbox{, if $z \le \sqrt{D_{n\ell}/n^4}$}\\
D_{n\ell}/(n^4 z^2) & \mbox{, if $z > \sqrt{D_{n\ell}/n^4}$}
\end{array} \right.
\end{equation} 
with $D_{n\ell} = 6 n^2 (n^2 - \ell^2 - \ell - 1)$. When used in Eq.~(\ref{integral}), this transition probability provides the Born approximation for the rate
\begin{equation}\label{born}
\tilde k^{(B)}(n,\ell,\theta) = \frac{1 - e^{D_{n\ell}\theta/2 n^4}}{2\theta} + \frac{D_{n\ell}}{4n^4} \Gamma(0, D_{n\ell}\theta/2n^4)
\end{equation}
where $\Gamma(0,x)$ is the incomplete gamma function. The PS64 result was derived under the same conditions with the additional tacit assumption that the cutoff impact parameter $R_c$ is greater than the $R_1$ parameter ($R_c > R_1$) for any projectile speed. Since the transition impact parameter increases with the speed of the projectile as $\sim 1/v$, the thermal average Eq.~(\ref{thermal}) will have a contribution from small speeds for which $R_1 > R_c$. This contribution, neglected in PS64, diminishes as $R_c \rightarrow \infty$.
Indeed, the PS64 rate is obtained from Eq.~(\ref{born}) in the $\theta \rightarrow 0$ limit
\begin{equation}\label{PS}
\tilde k^{(PS)}(n,\ell,\theta) = \frac{D_{n\ell}}{4n^4} \left[ 1 - \gamma - \log(D_{n\ell}\theta/2n^4)\right]
\end{equation}
where $\gamma$ is the Euler constant. This corresponds to Eq.(43) in PS64. The rate coefficients in Equation ~(\ref{PS}) become negative for $n$ large enough such that $D_{n\ell}\theta/n^4 > 2e^{1-\gamma}$. In other words, the PS64 approximation is limited to cases where 
$n^2 (n^2 - \ell^2 - \ell - 1) n_e/T^2 < 2.98\times 10^9$, which for small $\ell$ reduces roughly to $n^4 n_e/T^2 < 3\times 10^9$, and for large $\ell$ to $n^3 n_e/T^2 < 10^9$.

A semiclassical transition probability, compatible with the Born approximation and increasing linearly for small impact parameter, as predicted by the classical approximation, was obtained in (VOS17). 
In terms of the parameter $z$, the unresolved transition probability, summed over the $\ell'$ and slightly modified in order to give a better agreement with the quantum results, is
\begin{equation}
P^{(SC)} = \left\{ \begin{array}{ll}
z/\sqrt{6 D_{n\ell}/n^4}   & \mbox{, if $z \le \eta \sqrt{3 D_{n\ell}/2n^4}$}\\
3 j_1(\sqrt{3 D_{n\ell}/2n^4}/z)^2  & \mbox{, if $z > \eta \sqrt{3 D_{n\ell}/2n^4}$}
\end{array}\right.
\end{equation}
where $j_1(x)$ is the spherical Bessel function of order 1, and $\eta = 0.277855$ is the solution of the equation $j_1^2(1/x) = x/6$. When used in Eq.~(\ref{integral}), one gets the semiclassical approximation for the rate coefficient
\begin{equation}
\tilde k^{(SC)}(n,\ell,\theta) = \frac{D_{n\ell}}{4 n^4} \left[ 
\frac{3\sqrt{\pi}}{4 x^{3/2}} \mbox{erf}(\eta \sqrt{x}) - \frac{3\eta}{2x}\;e^{-\eta^2 x} + 
18 \int_0^\infty j_1^2(1/y)\; y e^{-x y^2}\; dy
\right]
\end{equation}
where $x = 3 D_{n\ell}\theta/4n^4$, and erf is the error function.  The integral can be treated as a Mellin-Barnes integral \citep{Paris2001} to obtain an asymptotic expansion in the parameter $x$, because
\begin{equation}\label{j1integral}
I(x) = 18 \int_0^\infty j_1^2(1/y)\; y e^{-x y^2}\; dy = \frac 1{2\pi i} \int_{c-i\infty}^{c + i\infty} F(1-s) G(s)\; ds
\end{equation}
where the integral goes along a line parallel with the imaginary axis with $0 < c < 2$, and where $F$ and $G$ are the Mellin transforms
\begin{equation}
F(1-s) = 18 \int_0^\infty j_1^2(1/y) \; y^{-s}\; dy = 18\;\frac{2^{3-s}\cos(\pi s/2) \Gamma(s-1)}{(s-6)(s-4)(s-3)}
\end{equation}
and
\begin{equation}
G(s) = \int_0^\infty  e^{-x y^2}\; y^{s-1}\; dy = \frac{\Gamma(s/2)}{2 x^{s/2}}
\end{equation}
Therefore the integral is
\begin{equation}
I(x) = 18 \int_{c-i\infty}^{c + i\infty} \frac{2^{2-s}\cos(\pi s/2) \Gamma(s-1)}{(s-6)(s-4)(s-3)} \frac{\Gamma(s/2)}{x^{s/2}}\; ds
\end{equation}
An expansion for small $x$ can be obtained by completing the integration along another line parallel with the imaginary axis, to create a contour around poles on the negative real axis and using the residue theorem. The poles of the integrand occur at negative even numbers $s = 0, -2, -4, \ldots, -2k, \ldots$ and are double. The double nature of the poles is the origin of the logarithm in the expansion of $I(x)$. The poles at $s = 1, -1, -3, \ldots$ are removable and do not contribute to the expansion. Therefore, after calculating the residues, one obtains 
\begin{equation}\label{expansion}
I(x)  = \sum_{k=0}^{N} \mbox{Res}(s = s_k) + {\cal O}(x^{N+1}) = \sum_{k=0}^{N} A_k (B_k - 3\gamma - 2\ln 2 - \ln x) x^k + {\cal O}(x^{N+1})
\end{equation}
where $\gamma$ is the Euler constant, and the coefficients are $A_k = 18 \times 4^k/(k! (2k+1)! (k+2)(k+3)(2 k + 3))$ and
$B_k = 1/(k+2) + 1/(k+3) + 2/(2k+3) + 1 + 1/2 + \ldots + 1/k + 2 + 2/3 + \ldots + 2/(2k+1)$.
With the first 11 terms listed in Table~\ref{table1} the expansion of Eq.~(\ref{expansion}) is accurate within one part in $10^4$ for $x  \le 50$.

\begin{table}
\caption{\label{table1} Coefficients $A_k$ and $B_k$ for the expansion of the integral in  Eq.(\ref{j1integral}).}
\begin{tabular}{lll}
k   & $A_k$                                      			& $B_k$        \\
\hline
0 	&1														&7/2 				= 3.5  \\
1	&1/5						= 0.2						&113/20				= 5.65 \\
2	&3/350						= 0.00857142857142857		&2857/420			= 6.80238095238095 \\
3	&2/14175					= 0.000141093474426808		&4793/630			= 7.60793650793651  \\
4	&1/873180					= 1.14523924047734$\times 10^{-6}$	&28526/3465			= 8.23261183261183 \\
5	&1/189189000				= 5.28571957143386$\times 10^{-9}$	&3151273/360360		= 8.74479131979132 \\
6	&1/65675610000				= 1.52263526749123$\times 10^{-11}$	&1102667/120120		= 9.17971195471195 \\
7	&1/34192364456250			= 2.92462956540933$\times 10^{-14}$	&2439746/255255		= 9.55807329924977 \\
8	&1/25408725942600000		= 3.93565581469558$\times 10^{-17}$	&575762023/58198140	= 9.89313443694249 \\
9	&1/25932145697017560000		= 3.85621772946852$\times 10^{-20}$	&42376261/4157010	= 10.1939280877361  \\
10	&1/35244143470037502000000	= 2.83735083773604$\times 10^{-23}$	&1000753049/95611230 = 10.4668985954893 \\
\hline
\end{tabular}
\end{table}

A modified PS64 approximation was obtained in \citep{Guzman2017} by replacing the original constant 1/2 for $b < R_1$ with a linearly increasing transition probability. When written in terms of the { dimensionless} parameter $z$, { as defined in the discussion following Eq. (\ref{rate})}, this approximate transition probability is
\begin{equation}
P^{(PS-M)} = \frac 12 \left\{ \begin{array}{ll}
z \; (2P_1)^{3/2} / \sqrt{D_{n\ell}/{n^4}}   & \mbox{, if $z \le \sqrt{D_{n\ell}/(2 P_1 n^4)}$}\\
D_{n\ell}/(n^4 z^2) & \mbox{, if $z > \sqrt{D_{n\ell}/(2 P_1 n^4)}$}
\end{array} \right.
\end{equation}
with the free parameter $P_1$ equal to the transition probability at the matching impact parameter. The integral
Eq.~(\ref{integral}) in this case yields the PS-M approximation 
\begin{equation}\label{PS-M}
\tilde k^{(PS-M)}(n, \ell, \theta) = \frac{D_{n\ell}}{4n^4} \left[ 
\frac{\sqrt{\pi}}{2 \beta^{3/2}} \;\mbox{erf}(\sqrt{\beta}) - \frac 1\beta e^{-\beta} + \Gamma(0, \beta) 
\right]
\end{equation}
with parameter $\beta = D_{n\ell}\theta/(4 P_1 n^4)$.

\end{document}